# FM Band Channel Measurements and Modeling

Omar Ahmadien*, Nann Win Moe Thet†, and Mehmet Kemal Ozdemir‡

*‡School of Engineering and Natural Sciences, Istanbul Medipol University, Istanbul, Turkey

† Advanced Wireless & Communication Research Center (AWCC), University of Electro-Communications, Tokyo, Japan

*omarahmadin@st.medipol.edu.tr, †nann@awcc.uec.ac.jp, ‡mkozdemir@medipol.edu.tr

*Abstract*—As FM coverage is so ubiquitous around the world, several applications can be considered to better exploit this useful band. Thus, it is of significant interest to investigate and characterize channel properties of the FM Band for the potential two-way digital wireless systems. In this paper, we present the results of field measurements at 86 MHz conducted at Gebze, Kocaeli, Turkey. Through the measurements, some of the FM channel characteristics are identified. Measurements are performed for urban, hilly terrain, and rural areas. Our results show that the FM channel expectedly has a large coverage area but at the same time, it possesses large channel excess delays. While most of COST-207 channel power delay profile models are also applicable for the FM Band, for urban environments and hilly terrain environments, channel clusters and excess delays are higher than those of COST-207 models. As 5G systems aim to utilize lower frequency bands for supplementary links, FM Band can be considered as one of the potential bands and the channel models proposed in this study can then be exploited for performance analysis.

## I. INTRODUCTION

Over the past decades, the behavior of wireless channels has been investigated and then has been standardized in order to better analyze various wireless systems. Wireless channel behavior at a certain frequency band can be better understood via real measurements in different types of environments of indoor or outdoor. Once the behavior of the wireless channel is determined, it can then be described statistically like the famous Saleh and Valenzuela model [1]. The statistical model may have some adaptive parameters so that a wider range of wireless channels can be supported, i.e., different frequencies or different environment types [2], [3]. Due to the limited resources of the allocated known frequency bands, researchers are moving towards 5G and are aiming to achieve higher data rates through more efficient utilization of the scarce frequency bands. On the other hand, certain frequency bands are poorly utilized, such as TV broadcast bands. These bands are now being considered either for two-way digital communication or for the supplementary links of 5G downlink channels [4].

Many studies for the white space TV bands were performed. One of the earliest studies was done in 2012 at Wilmington, NC, USA as the government connected some of the city's infrastructures, provided public wireless broadband to visitors in certain areas, and remotely monitored parking lots for security [5]. This initiative happened only after the FCC approved the first device to use the TV white space spectrum in 2011, which is made by Koos Technical Services (KTS). The device is called the Agility White Space Radio (AWR) [6], [7]. In 2012, the first patent to utilize the TV white space band was filed by [8], while in 2013, the IEEE 802.11af standard for wireless communication published White-Fi [9].

There have not been many studies that focus on channel measurement and modeling at FM band in the literature since FM band is generally considered for analog transmission only. In a study that focuses on FM band [3], the authors proposed a directional wireless channel model ideal for dense hilly terrain region at FM band based on COST 259 of 2 GHz [2]. Then, the proposed model was compared with adjacent FM band measurements [10] and ray tracing-based simulations.

Since there is limited literature on FM band studies that emphasize on some of the adjacent FM frequency bands are briefly discussed here. In studies [10], [11], the authors conducted channel measurement of the adjacent FM band at 145 MHz for hilly terrain regions. In work [2], the authors developed the COST 259 model for hilly terrain as well, but for the frequency range from 150 MHz to 2 GHz with parameterization at the frequencies between 500 MHz and 2 GHz. Another channel model for frequency bands from 150 MHz to 2 GHz is the COST 231 model and its extension, which was introduced in the study [12]. On the other hand, path loss models for 110.6 MHz and FM frequency bands are proposed in the work [13] and work [14], respectively. The COST-207 channel model that is based on the ultra-high frequency (UHF) band used for GSM between 8 and 10 MHz band was developed in the study [15].

### A. Motivation

Over time, the FM radio users are tending to move from traditional radio devices to online radio, especially the younger generation from 13 to 35 year-olds in the US [16] and the UK [17]. Moreover, between 2003 and 2013, there has been a steady increase in online radio revenue worldwide from 278 Million$ to 827 Million$, with an average annual growth rate of 28% per year [18]. This trend indicates that the traditional analog broadcast of FM radios will phase out in the near future. Hence, a utilization and standardization for this band is expected to happen, similar to what has happened in the analog TV band.

This project has received funding from the European Union's Horizon 2020 research and innovation programme under the Marie Skłodowska-curie grant agreement No 706929.

A lot of measurement campaign and statistical channel models with power delay profiles (PDPs) at millimeter-wave (mmWave) frequency bands and adjacent FM bands have been performed and proposed in the literature. However, due to the availability of FM band replaced by online radio services, there is a need to study and analyze channel modeling for future digital data transmission at FM bands. The band has good propagation properties such as higher penetration and longer range that make the band suitable for long range data communication. Therefore, in this paper, we present FM band channel measurements for the channel modeling, especially focusing on PDPs models in different environments.

*B. Contributions and paper organization*

As mentioned above, there has been many works focusing on adjacent FM bands measurements and PDP modeling, but, there are no studies and measurements specific for the FM band. To the best of our knowledge, this is the first extensive attempt for the understanding of the FM channel behavior. Field measurements in different scenarios are performed and multipath PDPs under different scenarios were obtained. Since earlier studies have performed path loss measurements, the measurement campaign of this study does not focus on the path loss, but rather focuses on the channel PDPs.

Firstly, in Section II, we briefly explain the wireless channel model. Then, the measurement setup such as measurement locations and the transmitted/received signals, applied in this study is discussed in Section III. In Section IV, data processing and PDP algorithm are presented. PDP measurement results and the proposed models are then evaluated in Section V, followed by the conclusion in Section VI.

## II. WIRELESS CHANNEL MODEL

A stochastic, linear, time-varying system is commonly used to model the wireless radio channel. This is also a good model for vehicular channels in which the signal suffers from scattering as a result of the presence of reflecting objects and scatterers in the channel environment. The random change in signal amplitude and phase of the different multipath components, caused by these reflection effects, is one of the factors that cause fluctuations in signal strength, thereby leading to small-scale fading, signal distortion, or both [19].

Moreover, small-scale variations of the mobile radio signals are directly related to the channel impulse response of the radio channel. Thus, for a time-variant CIR, the passband impulse response is defined as a finite series of attenuated, timedelayed, phase shifted duplicates of the transmitted signal [19], [20]. This passband impulse response of a multipath channel is described by Equation (1); where $a_i(t)$ represents the real amplitudes and $\tau_i(t)$ represents excess delays, for the $i^{th}$ multipath component at time $t$.

$$h(\tau,t) = \sum_i a_i(t)\delta(\tau - \tau_i(t)) \quad (1)$$

To effectively characterize a channel, we utilize certain channel metrics that are computed from the time-varying impulse response. Typically, the baseband sampled version of the impulse response is used for metric calculations. We denote $h_b[m,n]$ for the discrete equivalent of baseband channels from its passband channel $h(\tau,t)$ [21]. We use the time domain model and get the $a_i(t)$ and $\tau_i(t)$ for each channel realization at a given receiver location. We then average many of these realizations over a period of time to better estimate the delay profile. From the average impulse responses, we then obtain the channel power delay profile.

## III. MEASUREMENT SETUP

A measurement setup is constructed to effectively investigate the behavior of the channel through data captured in the field. Off-line analysis using MATLAB to obtain power delay profiles and Doppler spread is carried out.

The measurement setup is shown in Fig. 1. Here for multipath measurements, we use a single receiver antenna that was fixed over a SUV (Sport Utility Vehicle) as in Fig. 2b. Our transmitter is a R&S SMBV100A Victor Signal Generator. It is used for transmitting a designed signal with a certain power level and a specific center frequency, which is 86 MHz. At the receiver side, we have a R&S FSV Signal and Spectrum Analyzer, which receives this signal in a form of I/Q data format. Our transmitter antenna is an omni-directional FM dipole antenna with 2.15 dBi gain throughout the FM Band, while our receiver antenna is an outdoor telescopic FM antenna with 3 dBi gain. Both antennas are shown in Fig. 2.

*A. Measurement Place*

Our measurements took place at a site in Gebze, Turkey with different environments. With the SUV carrying the receiver set and a fixed transmitter on top of a building at Gebze Technical University, we were able to roam over different environments and scenarios for various channel types. For stationary measurements, we stopped and captured our data points at each measurement location, while for high speed measurements we took measurements continuously. The measurement points for one of the measurement sites that is in the range of 7 km is given in Fig. 3. On the other hand, a farther measurement site is depicted in Fig. 4.

The transmitter was on top of the Electrical Engineering building of Gebze Technical University, which has a height of 15 m, and the receiver was put on the roof of the SUV, whose height is approximately 2 m. We used Mini-Circuits LZY-1 amplifier, which is an Ultra-Linear RF amplifier with an output of up to 50 Watts for the frequency range of 20-512 MHz. The amplifier was hit with 0 dBm so that the antenna input power is 47 dBm.

For the synchronization between our transmitter and receiver, we used a GPS clock generating device, which is EPSILON CLOCK (MODEL EC1S) made by SPECTRA-COM [22]. It generates and distributes a highly accurate and stable frequency source disciplined using GPS signals as input. The device is able to generate a synchronized time reference.

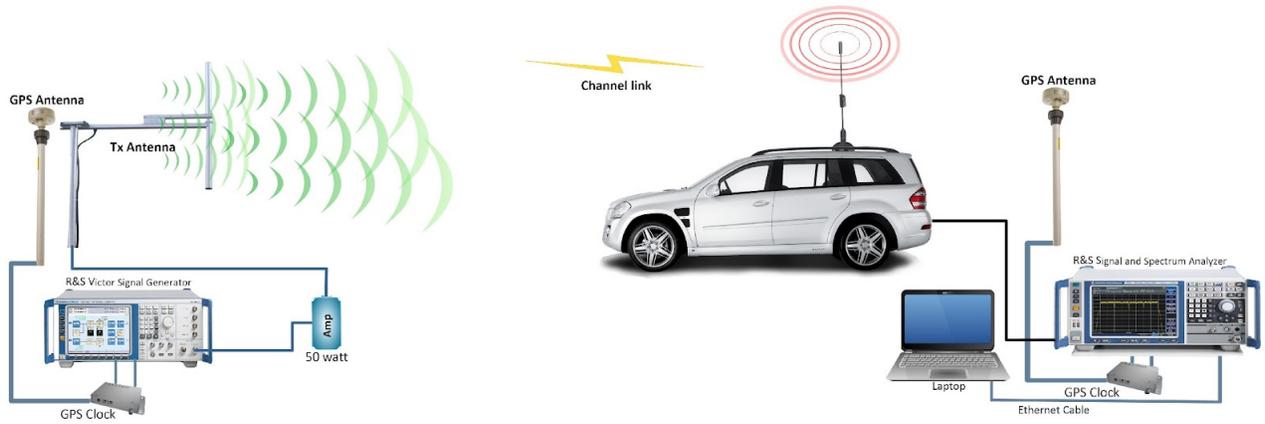

Fig. 1. Overall Measurement Setup

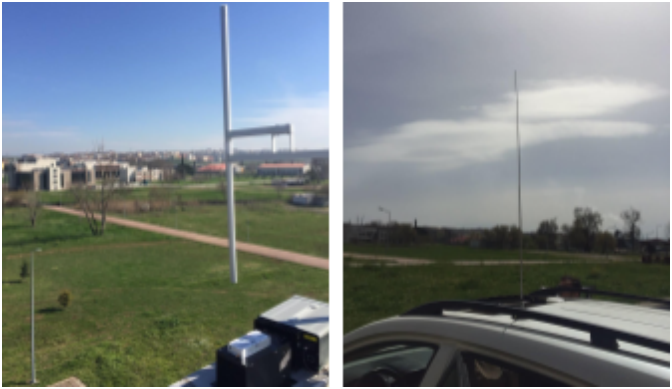

(a) Transmitter antenna and power amplifier.

(b) Receiver antenna on a SUV.

Fig. 2. Transmitter antenna with amplifier and receiver antenna.

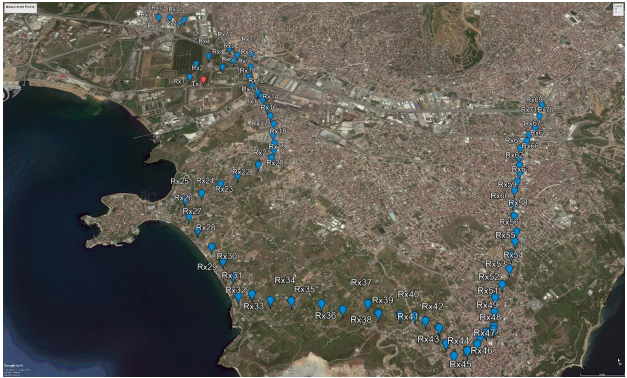

Fig. 3. Measurement Site 1: Blue points are the receiver points, while the transmitter is in red for near field, ranging up to 7 km.

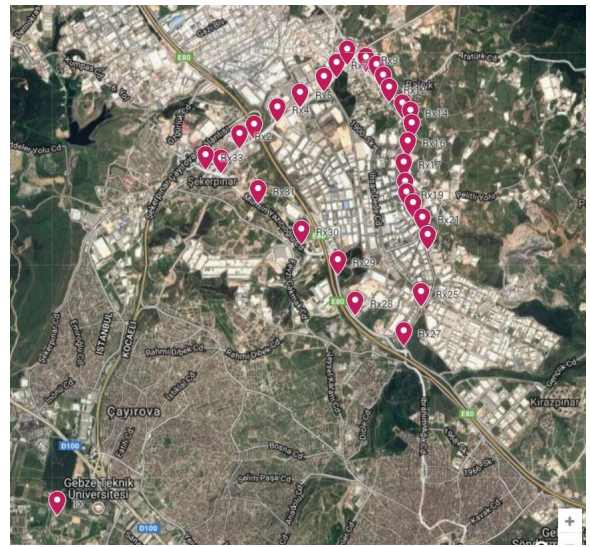

Fig. 4. Measurement Site 2: Urban area with more than 9 km away from the transmitter point.

The clock distributes a 1 PPS signal. It also generates a frequency reference at 10 MHz. In this paper, measurement results from the campaigns that are carried out during sunny days of April 2018 are presented.

### B. Transmitted and Received Signals

Our baseband transmitted signal is a typical maximum length (m-sequences) [20] of length $L = 2^m - 1$. We used a sequence-length of $L = 1023$ (with $m = 10$) and the sequence elements (chips) were transmitted at a rate of $1 \times 10^6$ Sample Rate/Hz, giving a chip-duration of $T_c = 1$ $\mu s$. The same sequence was padded with zeros so that a total length of 1100 samples is obtained. This sequence was then continuously repeated for $T_s = 1100$ $\mu s$. We generated the sequence by using MATLAB and converted it to ".wv" file format before uploading it to the transmitter, thanks to the R&S ARB Toolbox PLUS V 2.4 software.

In order to capture the received signal, we utilized the R&S IQ Wizard of R&S spectrum analyzer. The captured I/Q data is then imported to MATLAB for further processing and analysis.

## IV. DATA PROCESSING AND POWER DELAY PROFILES

Processing is done off-line by using the MATLAB environment. We have mainly focused on the generation of channel PDPs. Data format is ".mat" with I and Q separated. The data from the measurements campaign is available from the current research website [23]. The PDPs are obtained by estimating the channel impulse responses from the I/Q data. This is achieved through cross-correlation with the transmitted m-sequences that are available off-line. This principle of channel estimation by cross-correlation is described in Fig. 5 [20].

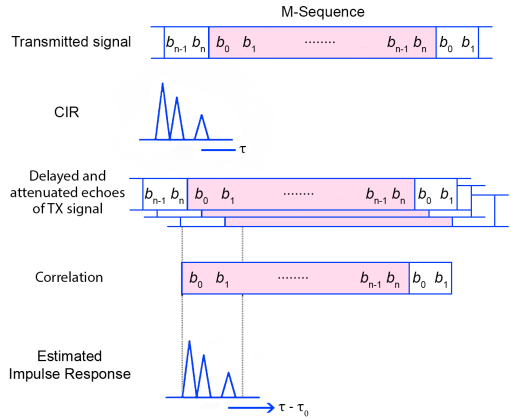

Fig. 5. The principle of channel estimation by cross-correlation.

The steps to obtain channel PDPs are explained in the Algorithm 1. The algorithm for the CIR averaging given in Algorithm. 1 averages the instantaneous CIRs from each msequence's cross-correlation. In a received sequence, typically there are more than 200 channel measurements. Mathematically, the averaging process can be described as

$$h_{PDP}[k] = \frac{1}{N_i} \sum_{A=0}^{N_i-1} |h_b[k, LA]| \quad (2)$$

where $L$ is the length of each sequence and $N_i$ is the number of averaged CIRs. Since the captured data sizes are set to be the same, $N_i$ is fixed in our measurements. The squared samples of the averaged channel impulse response then generates channel PDP.

---

**Algorithm 1** PDP Algorithm
1: **Inputs:** $Rx_{Data} = I + iQ$, $Tx_{Data} = M_{sequences}$ ($L = 1023$)
2: **Initialize:** $K = length(Rx_{Data})$,
3: $n = 1, 2, \ldots, \frac{K}{L}$ ;
4: **for** $i = 1 : K - L + 1$ **do**
5:    CIR(i) $= \sum Rx_{Data}(i : i+L-1) \circledast flip(Tx_{Data})$
6: **end for**
**return** $PDP = \frac{CIR(1)+CIR(2),\ldots,CIR(n)}{L}$

---

## V. RESULTS

Since the measurement sites included different types of environments, we have encountered different channel impulse responses. For the measurement campaign carried out in the site given in Fig.s4, we mostly obtained channel PDPs that either had a strong single multipath or another surviving multipath component nearby the first one. These two cases are shown in Fig.s 6 and 7. Hence, as we move away from the transmitter in urban environments, the clusters through which we receive the signals decrease. This agrees with the stochastic channel models developed for COST-259 [2] and for the FM Band channel models [3].

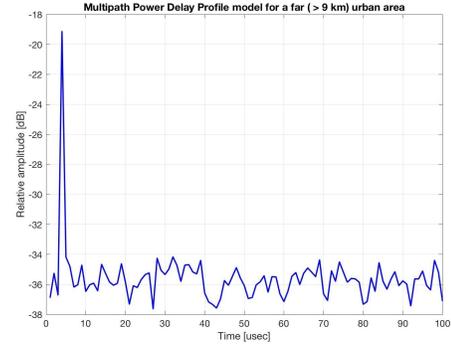

Fig. 6. PDP for site in Fig. 4 receiver location 28.

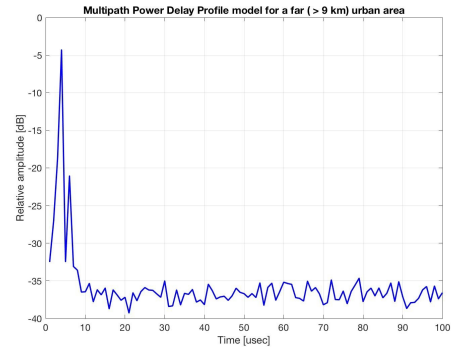

Fig. 7. PDP for site in Fig. 4 receiver location 5.

As for the site given in Fig. 3, we have obtained different channel PDPs for different environments. For the case of a rural environment, we have obtained a strong line-of-sight component as given in Fig. 8. When the measurement points were chosen from the urban areas, we started receiving more clusters and consequently we observed larger channel excess delays. These are shown in Fig.s 9a and 9b. For the case of Fig. 9b, we have observed channel excess delays as large as 30 $\mu s$, though the last pack of multipaths were 25 dB lower than the strongest path. Although the PDPs obtained through this study are in good agreement with COST-207 channel model [15], for bad urban cases, we have obtained channels with more clusters and larger excess delays. Hence, for the FM band with a bad urban environment, we recommend the channel PDP model given in Fig. 10. Assuming white sense stationary (WSS) channel, the PDP is theoretically calculated based on the square of amplitude of the channel with $L$ clusters as

$$PDP(\tau) = |h(\tau)|^2 = \sum_{i=1}^{L} |\alpha_i|^2 \delta(\tau - \tau_i), \quad (3)$$

where the power in dB scale at delay $\tau_i$ can be approximated using piecewise function

$$\alpha_i = \begin{cases} -1.7\tau_{k1} & \text{if } 0 \leq \tau \leq 10 \\ -1.76\tau_{k2} + 11.6 & \text{if } 10 \leq \tau \leq 35 \\ 55(0.85)^{\tau_{k3}} - 78 & \text{if } 35 \leq \tau \leq \infty \end{cases} \quad (4)$$

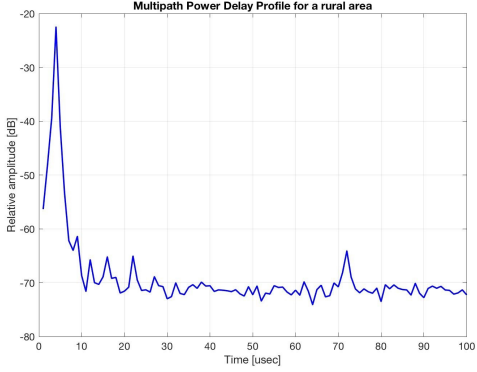

Fig. 8. PDP for site in Fig. 3 receiver location 25.

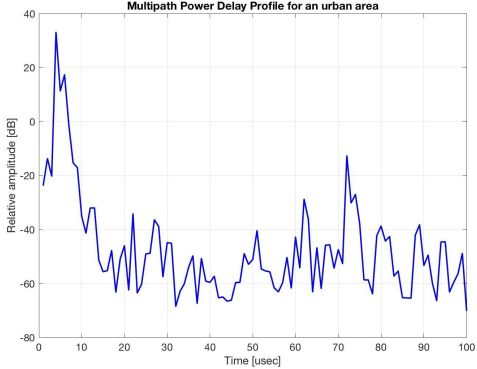

(a) At location 74.

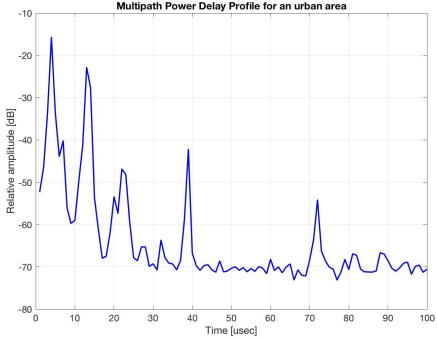

(b) At location 71.

Fig. 9. PDP for site in Fig. 3 receiver at (a) location 74 and (b) location 71

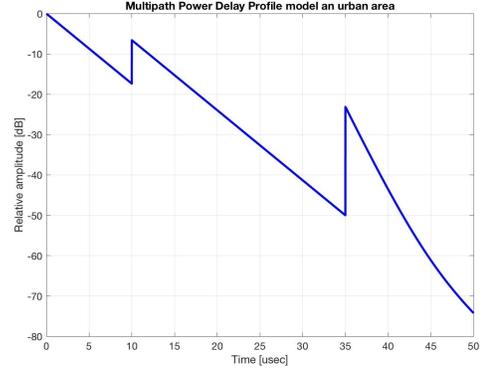

Fig. 10. FM band channel PDP model for bad urban environment.

For the hilly terrain regions, we have picked two receiver locations that present the typical behavior. Fig.s 11a and 11b show the channel PDP for these two cases. As can be observed, for the hilly terrain regions, we typically get multipath components and channel excess delays beyond 10 $\mu s$. Although most of the hilly terrain environment measurements are in agreement with COST-207 hilly terrain channel models [15], for some cases we have observed additional channel clusters. In order to model this behavior, we propose the channel PDP model given in Fig. 12. Similar to the bad urban case, the proposed PDP model can also be approximately expressed based on (3) with the power $\alpha_i$, given as

$$\alpha_i = \begin{cases} -8.6667\tau_{k1}, & \text{if } 0 \leq \tau_{k1} \leq 3 \\ -4.8684\tau_{k2} + 2.6053, & \text{if } 3 \leq \tau_{k2} \leq 6.8 \\ -4.2857\tau_{k3} + 31.6429, & \text{if } 11 \leq \tau_{k3} \leq 14.5 \\ -30.5 & \text{else} \end{cases} \quad (5)$$

We have also measured the Doppler spectrum of the FM Band but since the band has a relatively low frequency, the Doppler shift is in Hertz units and therefore channel coherence time is very large compared to the typical frame length of digital communication systems. Hence, the band can be effectively used for high-speed mobiles.

## VI. CONCLUSION

In this paper, we have presented the measurement results for FM Band so that better channel models can be developed. We have observed that the FM channel possesses large delay spread especially in bad urban environments, making 4G and 5G's main waveform orthogonal frequency division multiplexing (OFDM) transmission less efficient. However, the band has good penetration through different environments, and hence, with new waveform design in 5G, like non-orthogonal multiple access (NOMA), it can be utilized for supplementary downlink channels, instead of being under-utilized for analog FM music broadcast. Moreover, with 5G systems also targeting vehicle-to-vehicle communication, since the band has large channel coherence time, it stands as a strong candidate to be used as the frequency band of vehicle-to-vehicle communication.

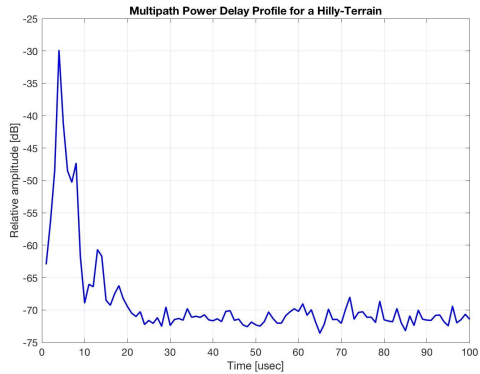

(a) At location 36.

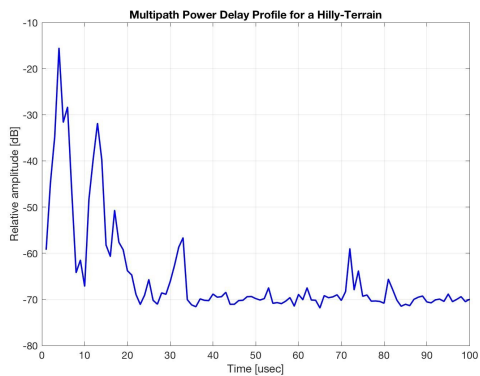

(b) At location 37.

Fig. 11. PDP for site in Fig.3 receiver at (a) location 36 and (b) location 37

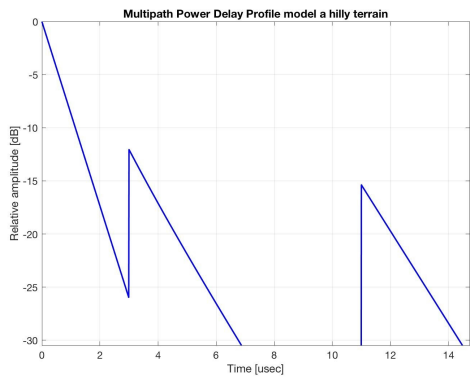

Fig. 12. FM band channel PDP model for a multi-cluster hilly terrain environment.

## REFERENCES

[1] A. Saleh and R. Valenzuela, "A statistical model for indoor multipath propagation," *IEEE J Sel Top Signal Process.*, vol. 5, no. 2, pp. 128–137, February 1987.

[2] H. Asplund, A. Glazunov, A. Molisch, K. Pedersen, and M. Steinbauer, "The cost 259 directional channel model-part ii: macrocells," *IEEE Trans Wirel Commun.*, vol. 5, no. 12, pp. 3434–3450, 2006.

[3] A. H. Mohammed and M. K. Ozdemir, "A directional fm channel model for contemporary wireless systems," *Can. J. Electr. Comput. Eng.*, vol. 39, no. 4, pp. 311–321, 2016.

[4] "Federal Communications Commission," http://www.fcc.gov, . Published May, 2018.

[5] E. Woyke, "World's first commercial white spaces network launching today in north carolina," https://www.forbes.com/sites/elizabethwoyke/2012/01/26/worlds-first-commercial-white-spaces-network-launchingto day-in-north-carolina, . Published January, 2012.

[6] "Agility white space radio (AWR)," http://www.ktswireless.com/wpcontent/uploads/2013/11/AWR-Quick-Start-Manual.pdf.

[7] D. Dante, "FCC approves first white space device and spectrum database," https://www.theverge.com/2011/12/22/2656177/fccapproves-first-white-space-spectrum-device-database, Dec . Published December, 2011.

[8] A. Hassan, D. Reed, P. Garnett, and B. Anders, "White space utilization," https://patents.google.com/patent/WO2014042964A1/tr, . Published June, 2012.

[9] A. Flores, R. Guerra, E. Knightly, P. Ecclesine, and S. Pandey, "IEEE 802.11af: a standard for TV white space spectrum sharing," *IEEE Commun. Mag.*, vol. 51, no. 10, p. 92–100, 2013.

[10] A. Aitalieva, "Vhf channel modeling for wireless sensor networks," Master's thesis, . Gebze Technical University, Gebze, Turkey, ; 2017.

[11] A. Aitelieva, G. Celik, and H. Celebi, "Ray tracing-based channel modelling for vhf frequency band," ser. 2015 23nd Signal Processing and Communications Applications Conference (SIU). IEEE, 2015, pp. 1385–1388.

[12] G. Pedersen, "Cost 231-digital mobile radio towards future generation systems," ser. Cost 231-Digital Mobile Radio Towards Future Generation Systems. EU, 1999, pp. 92–96.

[13] M. Meeks, "Vhf propagation over hilly, forested terrain," *IEEE Trans Antennas Propag.*, vol. 31, no. 3, pp. 483–489, 1983.

[14] P. Pathania, P. Kumar, and S. B. Rana, "A modified formulation of path loss models for broadcasting applications," *Int. J. Recent Technol. and Eng*, vol. 3, no. 3, pp. 44–54, 2014.

[15] C. of the European Communities, *Digital Land Mobile Radio Communications - COST 207: Final Report*. Official Publications of the European Communities, 1989.

[16] L. Indvik, "Study: More young adults tuning out terrestrial radio while driving," http://mashable.com/2013/04/02/internet-radio-study-npd, Apr . Published April, 2013.

[17] "Half of UK listens to online radio," http://advanced-television.com/2013/06/12/half-of-uk-listens-to-online-radio, . Published June, 2013.

[18] "Revenue online radio 003-2013 forecast," https://www.statista.com/statistics/265681/online-radio-revenue-worldwide.

[19] T. Rappaport, *Wireless Communications: Principles and Practice*, 2nd ed. Upper Saddle River, NJ, USA: Prentice Hall PTR, 2001.

[20] A. Molisch, *Wireless communications*. Oxford, UK: Wiley-Blackwell, 2010.

[21] D. Tse and P. Viswanath, *Fundamentals of wireless communication*. Cambridge: Cambridge University Press, 2008.

[22] "Epsilon clock model EC1S user's manual," https://spectracom.com/documents/users-manual-epsilon-clock-model-ec1s.

[23] "FM for next generation research group (FM-4NXTG) Dataset," https://github.com/NanntThet/FM-4NXTG.